\documentclass[10pt,twocolumn]{article}
\usepackage[utf8]{inputenc}
\usepackage{amsmath}
\usepackage{graphicx}

\title{Estimating the age of renal tumors} \author{Allen
  B. Downey\footnote{Professor of Computer Science, Olin College of
    Engineering, Needham MA 02492.  Email: {\tt
      allen.downey@olin.edu}, Web: {\tt allendowney.com}}} \date{March
  2012}

\begin{document}

\maketitle 

\begin{abstract}
We present a Bayesian method for estimating the age of a renal tumor
given its size.  We use a model of tumor growth based on
published data from observations of untreated tumors.  We find, for
example, that the median age of a 5 cm tumor is 20 years, with
interquartile range 16--23 and 90\% confidence interval 11--30 years.
\end{abstract}

\section{Introduction}
  
For some cancer patients it is important to estimate a tumor's date of
formation; for example, benefits provided by the U.S. Department of
Veterans Affairs depend on whether it is likely that a tumor formed
while the patient was in military service, among other considerations.
Doctors are currently unable to provide a statistical estimate of when a
tumor formed.

For renal cancers, we have reliable measurements for the rate of growth
during a period of observation; this paper presents a method to use
this data to estimate the distribution of ages for a renal tumor based
on size at diagnosis.

\subsection{Prior work}

\begin{figure}
\centerline{\includegraphics[width=3.0in]{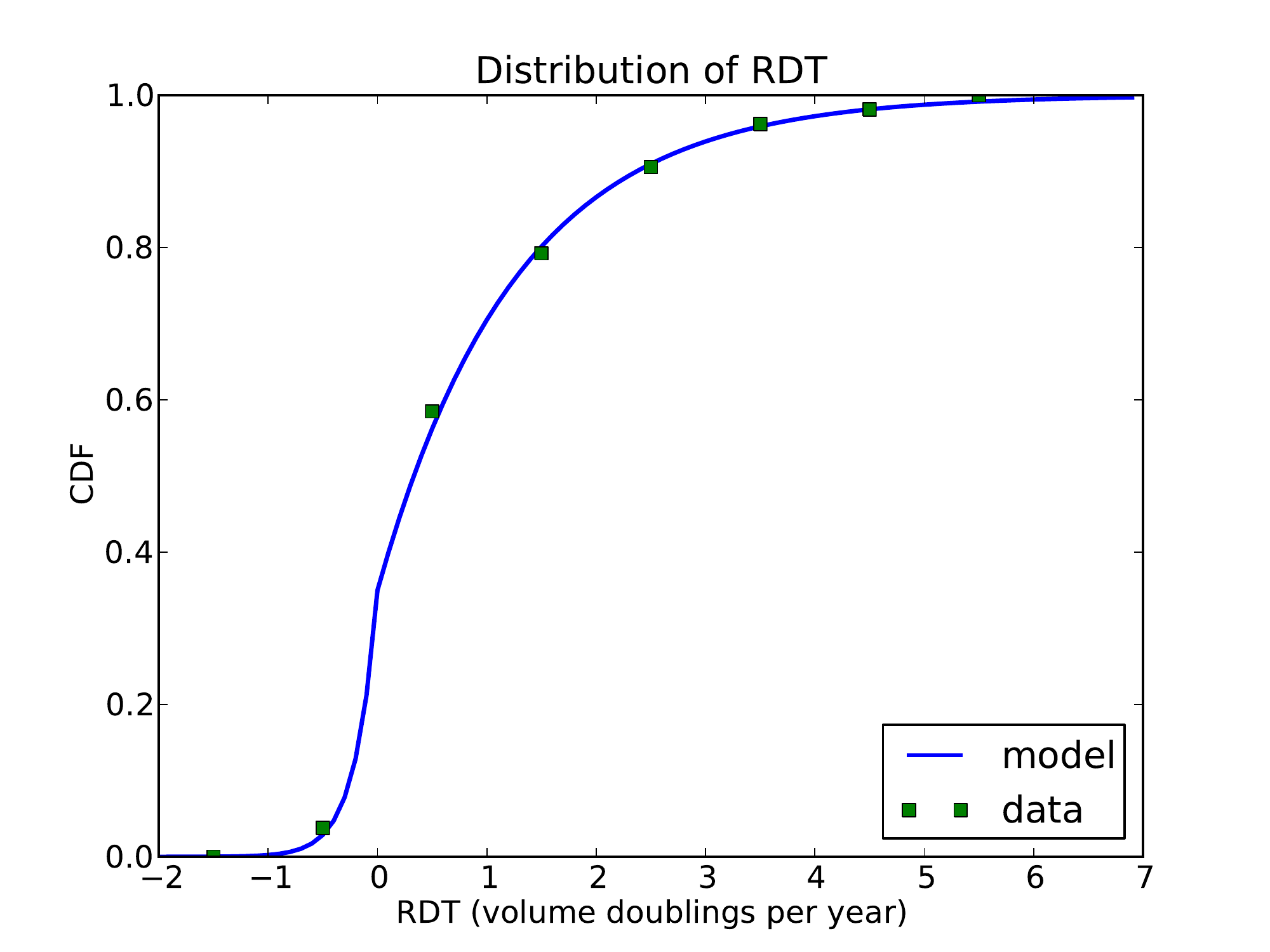}}
\caption{Distribution of RDT from Zhang et al. }
\end{figure}

Several studies report growth rates for patients with untreated
renal tumors [1-11].  We use this data to estimate the distribution
of growth rates that is the basis of our model.

Some studies report that different subtypes grow at different rates,
but Zhang et al [11] conclude ``Growth rates in renal tumors of
different sizes, subtypes and grades represent a wide range and
overlap substantially.''  Other studies are consistent with this
conclusion, so our growth model does not take into account tumor
subtype or grade.

\section{Methodology}
  
We use data from Zhang et al [11] to estimate the distribution of
growth rates during a period of observation.  We use this distribution
to generate tumor growth histories.  These simulations yield $P(d|t)$,
the distribution of diameter, $d$, as a function of the age of the
tumor, $t$.  Then we compute $P(t|d)$, the distribution of age as a
function of size.
 
\subsection{Distribution of growth rates} 

Zhang et al report reciprocal doubling times (RDT), in units of
doublings per year, for 53 ``patients who underwent nephrectomy from
1989 to 2006 who did not receive preoperative chemotherapy or
radiation therapy and underwent at least two preoperative contrast
material-enhanced CT examinations (at least 3 months apart)''.

We use data from their Figure 3 to construct the cumulative
distribution function (CDF) of RDT, shown in Figure 1.
The line shows a model of the data as a mixture of exponential
distributions: with probability $p = 0.35$ we generate a negative RDT
with $\lambda_2 = 5.0$; otherwise we generate a positive RDT with
$\lambda_1 = 0.79$.  The model fits the data well, allowing us to
interpolate between data points and characterize tail behavior.

\subsection{Simulated growth histories} 

\begin{figure}
\centerline{\includegraphics[width=3.0in]{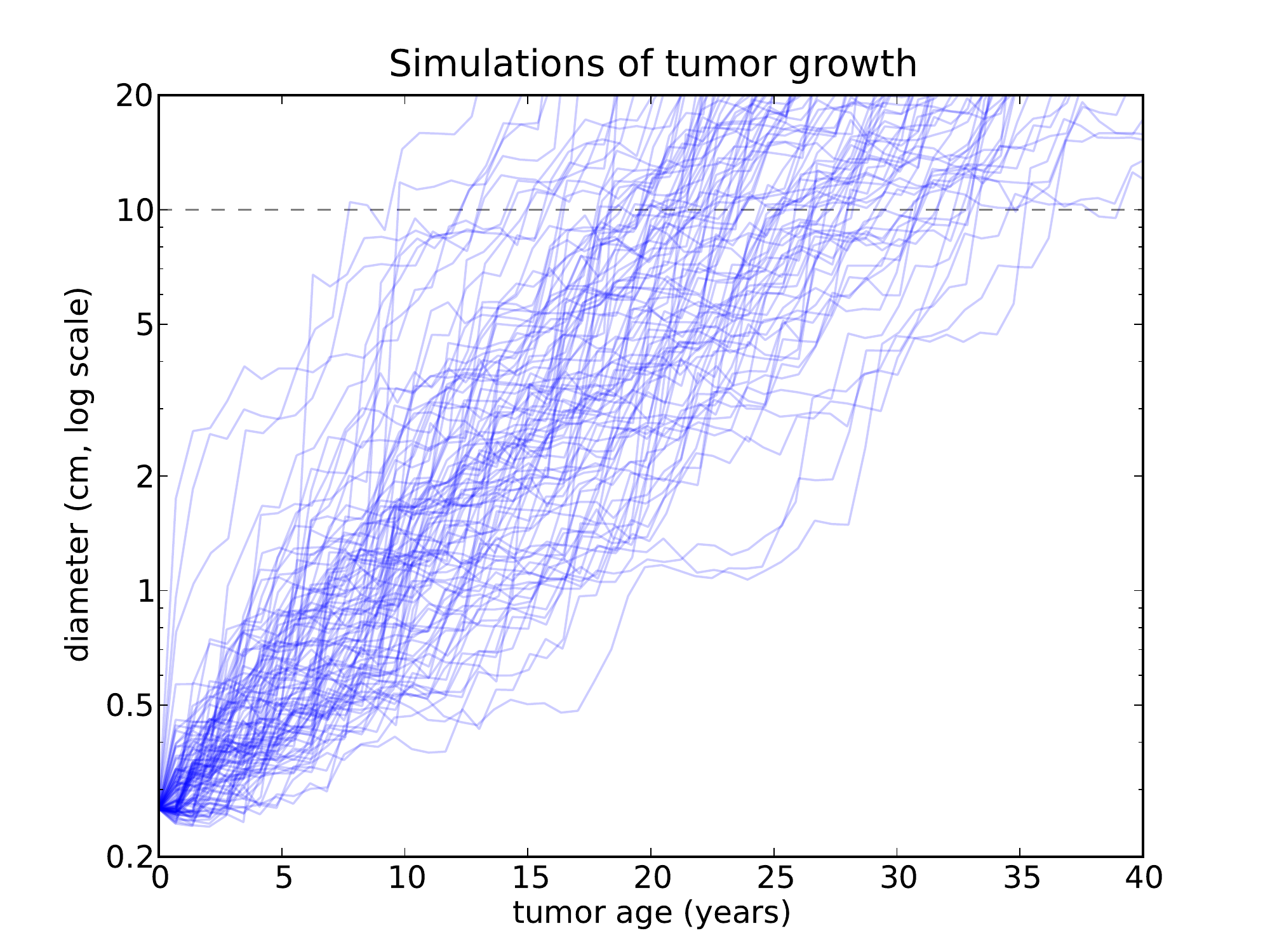}}
\caption{100 simulations of tumor growth. }
\end{figure}

To simulate tumor growth, we start at a hypothetical time, $t_0$, when
tumor volume is $V_0 = 0.01$ mL (diameter 0.27 cm), and repeat these
steps:

\begin{enumerate}

\item Choose a random value from the modeled distribution of RDT.

\item Compute the volume at the end of the interval, $V_{i+1} =
  2^{h \cdot RDT} V_i$, where $h$ is the duration of the interval in years.
In Zhang et al, the median time between initial and final scans
is 245 days, so we used this value for the interval, $h$. 

\item Repeat until volume exceeds $V_{max} = 4200$ mL (diameter 20 cm).

\end{enumerate}

Figure 2 shows the result of 100 simulations.  The vertical scale is
$d$, diameter in cm.  Using these simulations, we compute $P(d|t)$ the
distribution of size given $t$, time since $t_0$.

\subsection{Serial correlation}

\begin{figure}
\centerline{\includegraphics[width=3.0in]{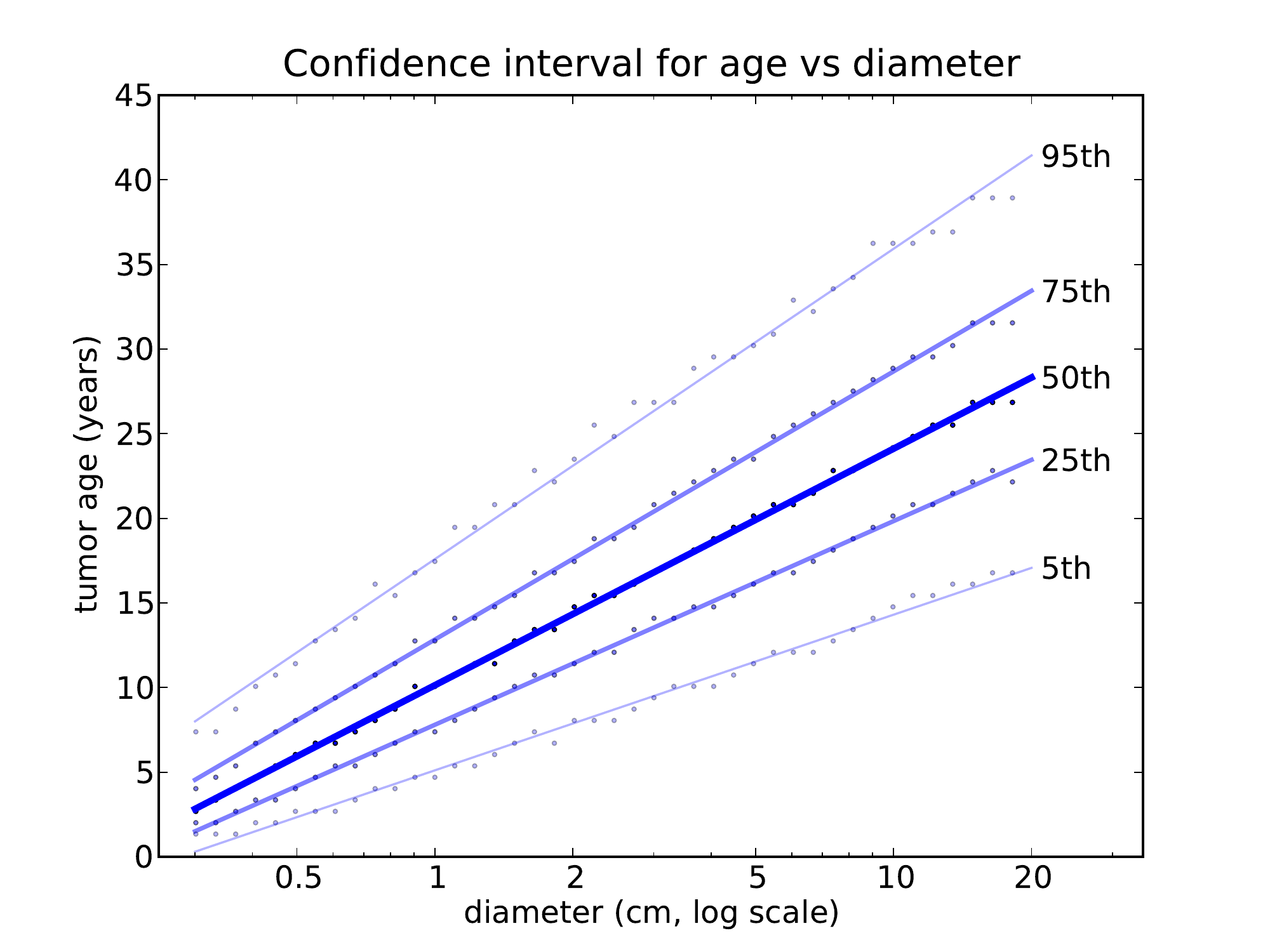}}
\caption{Percentiles of time since $t_0$. }
\end{figure}

\begin{table}
\begin{tabular}{|r||r|r|r|r|r|}
\hline
Diameter   & \multicolumn{5}{c|}{Percentiles of age} \\
(cm)   & 5th & 25th & 50th & 75th & 95th \\
\hline
0.3 & 1.3  & 2.0  & 2.7  & 4.0  & 7.4 \\
0.4 & 2.0  & 3.4  & 4.7  & 6.7  & 10.1 \\
0.5 & 2.7  & 4.7  & 6.7  & 8.7  & 12.8 \\
0.7 & 4.0  & 6.0  & 8.1  & 10.7  & 16.1 \\
1.0 & 4.7  & 7.4  & 10.1  & 12.8  & 17.5 \\
1.3 & 6.0  & 9.4  & 11.4  & 14.8  & 20.8 \\
1.8 & 6.7  & 10.7  & 13.4  & 16.8  & 22.2 \\
2.5 & 8.1  & 12.1  & 15.4  & 18.8  & 24.8 \\
3.3 & 10.1  & 14.1  & 17.5  & 21.5  & 26.8 \\
4.5 & 10.7  & 15.4  & 19.5  & 23.5  & 29.5 \\
6.0 & 12.1  & 16.8  & 20.8  & 25.5  & 32.9 \\
8.2 & 13.4  & 18.8  & 22.8  & 27.5  & 34.2 \\
11.0 & 15.4  & 20.8  & 24.8  & 29.5  & 36.2 \\
14.9 & 16.1  & 22.2  & 26.8  & 31.5  & 38.9 \\
\hline
\end{tabular}

\caption{Percentiles of time since $t_0$.}
\end{table}

In our model of tumor growth, the growth rate during each interval is
independent of previous growth rates. 
It is plausible that, in reality, tumors that have grown quickly in
the past are more likely to grow quickly.

If this correlation exists, it affects the location and spread of our
results.  For example, running simulations with $\rho = 0.4$ increases
the estimated median age by about a year, and the interquartile range
by about 3 years.

However, if there were a strong serial correlation in growth rate,
there would also be a correlation between tumor volume and growth
rate, and prior work has shown no such relationship [7] [11].

There could still be a weak serial correlation, but since there is
currently no evidence for it, we report results based on simulations
with $\rho = 0$.


\subsection{Inverse probabilities}
  
In Figure 2 the dashed line at 10 cm illustrates our method for
inverting $P(d|t)$.  When a simulated history crosses
this line, that represents a point in time when a tumor could be
observed at this size.  Since we know the duration of each history, we
can construct the distribution of ages for a tumor observed at this
size.  Repeating this computation for each size, we construct $P(t|d)$.

\section{Results}

Figure 3 shows the confidence interval for $t$ as a function of size.
The points are data from simulation, which produces some variability due
to discrete approximation.  The lines are fitted to the data.  For each size,
we compute the median of $t$, interquartile range, and 90\% confidence
interval.  Table 1 shows these results numerically, for selected sizes.

\section{Conclusion}

Using Table 1 we can look up the size of a tumor and find the
distribution of time since $t_0$, which is a lower bound on the tumor's age.
This information is potentially useful to patients, doctors and insurers.

\section{References}

\noindent [1] Birnbaum BA, Bosniak MA, Megibow AJ, Lubat E, Gordon
RB. Observations on the growth of renal neoplasms. {\em Radiology}
1990;176:695–701.

\noindent [2] Bosniak MA. Observation of small incidentally detected renal
masses. {\em Semin Urol Oncol} 1995;13:267–272.

\noindent [3] Bosniak MA, Birnbaum BA, Krinsky GA, Waisman J. Small renal
parenchymal neoplasms: further observations on growth. {\em Radiology}
1995;197:589–597.

\noindent [4] Kassouf W, Aprikian AG, Laplante M, Tanguay S. Natural history of
renal masses followed expectantly. {\em J Urol} 2004;171:111–113.

\noindent [5] Fujimoto N, Sugita A, Terasawa Y, Kato M. Observations on the
growth rate of renal cell carcinoma. {\em Int J Urol} 1995;2:71–76.

\noindent [6] Oda T, Miyao N, Takahashi A, et al. Growth rates of primary and
metastatic lesions of renal cell carcinoma. {\em Int J Urol} 2001;8:473–477.

\noindent [7] Ozono S, Miyao N, Igarashi T, et al. Tumor doubling time of renal
cell carcinoma measured by CT: collaboration of Japanese Society of
Renal Cancer. {\em Jpn J Clin Oncol} 2004;34:82–85.

\noindent [8] Rendon RA, Stanietzky N, Panzarella T, et al. The natural history
of small renal masses. {\em J Urol} 2000;164:1143–1147.

\noindent [9] Sowery RD, Siemens DR. Growth characteristics of renal cortical
tumors in patients managed by watchful waiting. {\em Can J Urol}
2004;11:2407–2410.

\noindent [10] Volpe A, Panzarella T, Rendon RA, Haider MA, Kondylis FI, Jewett
MA. The natural history of incidentally detected small renal
masses. {\em Cancer} 2004;100:738–745.

\noindent [11] Zhang et al., Distribution of Renal Tumor Growth Rates Determined
by Using Serial Volumetric CT Measurements, January 2009 {\em Radiology},
250, 137-144.

\end{document}